\documentclass[twocolumn,aps,superscriptaddress,showpacs,nofootinbib,floatfix]{revtex4}
\usepackage{epsfig,bm,feynmf}
\usepackage{graphics}
\usepackage{amsmath}
\usepackage[normalem]{ulem}  
\usepackage[dvips]{color} 

\renewcommand\sout{\bgroup \color{red} \ULdepth=-.5ex \ULset}
\begin{document}


\title{$D$ meson mass and heavy quark potential at finite temperature}


\author{Philipp Gubler}%
\email{gubler@post.j-parc.jp}
\affiliation{Advanced Science Research Center, Japan Atomic Energy Agency, Tokai, Ibaraki 319-1195, Japan}

\author{Taesoo Song}\email{t.song@gsi.de}
\affiliation{GSI Helmholtzzentrum f\"{u}r Schwerionenforschung GmbH, Planckstrasse 1, 64291 Darmstadt, Germany}

\author{Su Houng Lee}%
\email{suhoung@yonsei.ac.kr}
\affiliation{Department of Physics and Institute of Physics and Applied Physics, Yonsei University, Seoul 03722, Korea}


\begin{abstract}
Based on the observation that the heavy quark - anti-quark potential value at infinity corresponds to twice the D meson mass, we constrain the
asymptotic value of the heavy quark potential in a hot medium through a QCD sum rule calculation of the D meson at finite temperature.
We find that to correctly reproduce the QCD sum rule results as well as a recent model calculation for the D meson mass near the critical temperature, the heavy quark potential should be
composed mostly of the free energy with an addition of a small but nontrivial fraction of the internal energy.
Combined with a previous study comparing
potential model results for the $J/\psi$ to a QCD sum rule calculation, we conclude that the composition of the effective heavy quark potential should depend on
the interquark distance. Namely, the potential is dominated by the free energy at short distance,
while at larger separation, it has a fraction of about 20\% of internal energy.
\end{abstract}


\maketitle

\section{Introduction}

Quarkonium is a colorless and flavorless bound state of a heavy quark and its anti-quark.
Since Matsui and Satz proposed the suppression of quarkonium in relativistic heavy-ion collisions as a signature of the quark-gluon plasma formation due to color screening~\cite{Matsui:1986dk},
this phenomenon has been investigated in numerous theoretical and experimental studies (see, for instance, the recent review in Ref.\,\cite{Rothkopf:2019ipj}).
One of the most critical quantities affecting quarkonium suppression is the potential between a heavy quark and anti-quark.
In lattice quantum chromodynamics (QCD), the free energy of the static quark antiquark pair can be obtained through the rectangular Polyakov loop along the space-time direction.
As one increases the temperature, the long distance part of the  free energy exhibits a sudden saturation near the critical temperature, which can be interpreted as the onset of deconfinement.
If the free energy is adopted for the potential of the heavy quark system \cite{Satz:2015jsa}, this behaviour leads to a weakening of the quarkonium binding with increasing temperature
so that the $J/\psi$, 1S bound state of a charm and anti-charm quark, dissolves around 1.1 $T_c$~\cite{Lee:2013dca}.
On the other hand, at finite temperature the heavy quark and anti-quark can themselves be polarized acquiring additional energy. This leads to the internal energy of the system as
another candidate for the heavy quark potential, which has an additional contribution from the entropy density~\cite{Kaczmarek:2002mc},
\begin{eqnarray}
U=F+TS,
\end{eqnarray}
where $F$, $T$ and $S$ are the free energy, temperature and the entropy density $S=-\partial F/\partial T$, respectively.
While the free energy  at large separation drops rapidly near the critical temperature, the entropy density becomes large, thus acting as a high potential wall, which prevents the quarkonium from dissolving.
As a result, if the internal energy is adopted for the heavy quark potential, $J/\psi$ binding persists beyond 1.6 $T_c$~\cite{Satz:2005hx}.
The free energy is a thermodynamical potential of the system which is surrounded by a thermal heat reservoir, such that the temperature is kept fixed by exchanging energy with that reservoir.
It measures the amount of available energy of the separated heavy quark anti-quark pair compared to that at different separation distance.
On the other hand, the internal energy is a thermodynamical potential of an isolated system so that the energy needed to polarize the heavy quark states should be added.
The question, which
thermodynamical potential is appropriate as a heavy quark potential at finite temperature, has been
controversially discussed in the literature~\cite{Wong:2004zr,Brambilla:2008cx}.

Several years ago, two of the present authors have calculated the strength of the $J/\psi$ wavefunction at the origin as a function of temperature by using
a QCD sum rule approach with gluon condensates reliably determined from lattice QCD calculations. The result was then compared with the $J/\psi$  wavefunction obtained by solving the Schr\"{o}dinger equation
either with the free energy or the internal energy as a heavy quark potential~\cite{Lee:2013dca}.
It was thus found that the strength of the $J/\psi$ wavefunction as well as the $J/\psi$ mass as a function of temperature obtained from the QCD sum rule calculation follow those obtained from
using the free energy as a potential. Meanwhile, the solutions from the internal energy potential significantly overestimate both quantities.
This comparison led to the conclusion that the free energy is the appropriate choice for the heavy quark potential at finite temperature rather than the internal energy, which is consistent with recent lattice
results extracted from the spectral functions of Wilson line correlators~\cite{Petreczky:2018xuh}.

In the potential model, not only the mass of the  $J/\psi$, but also that of the $D$ meson is closely related to the heavy quark potential.
This can be understood by considering the eigenenergy of an infinitely separated static charm quark pair, which corresponds to twice the $D$ meson mass, as the charmonium will be separated into
$D$ and $\bar{D}$ mesons at a large distance with vanishing interaction.
The $D$ meson mass at finite temperature hence depends on the behavior of the heavy quark potential at large distance, which is quite different for the free energy and internal energy cases near $T_c$.

In the past, several theoretical studies on the $D$ meson spectrum in a hot or dense medium and its application to heavy-ion collisions have been carried out (see, for example,
Refs.~\cite{Hayashigaki:2000es,Cassing:2000vx,Mishra:2003se,Tolos:2007vh,Suzuki:2015est,Carames:2016qhr}).
In QCD sum rules, the real part of the current correlator in the $D$ meson channel is related to its imaginary part through a dispersion relation.
The former is in the deep Euclidean region computed using the operator product expansion, which leads to an analytic expression with QCD condensates and corresponding Wilson coefficients.
The latter imaginary part is expressed as a spectral function composed of physical states with the same quantum number as the D meson.
To analyze the sum rules, one then assumes the spectral function to have the simple form of a single ground state and a smooth continuum.
The mass of the ground state is then extracted by matching the spectral function to the OPE result. It is thus possible to relate the temperature dependence of the D meson mass to that of the QCD condensates.

The main goal of this paper is to determine what kind of heavy quark potential can model the $D$ meson mass temperature dependence extracted from the QCD sum rule approach.
We find that  the heavy quark potential at large separation should be composed mostly of the free energy with an addition of a small but nontrivial fraction of the internal energy.
Combined with our previous results for the heavy quark potential obtained by comparing the potential model result for the $J/\psi$ to the QCD sum rule calculation, we
conclude that the effective heavy quark potential is dominated by the free energy at short distance, while at larger separation, it has a non-negligible fraction of the internal energy.

This paper is organized as follows.
In Section~\ref{potentialm} we introduce the heavy quark potentials from lattice QCD calculations and
show the obtained $D$ meson mass as a function of temperature for each of the various potentials.
The same $D$ meson mass is calculated from QCD sum rules in Section~\ref{QSR} and compared to the potential model results in Section~\ref{comparison}. The conclusion is given in Section~\ref{conclusion}.

\section{$D$ meson mass from heavy quark potential}\label{potentialm}

The strong interaction between the heavy quark and anti-quark is in our approach modeled as a combination of a Coulomb and linear potential.
The former is dominant at short, the latter at long distances. The linear potential is furthermore generally responsible for the confinement of hadrons.

\begin{figure}
\centerline{
\includegraphics[width=8.6 cm]{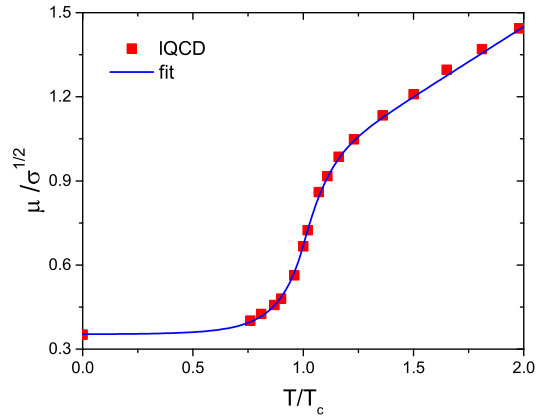}}
\caption{The screening mass $\mu$ scaled by $\sqrt{\sigma}$, extracted from lattice QCD
calculations~\cite{Satz:2005hx,Kaczmarek:1900zz,Digal:2005ht} (red points) and fitted to Eq.\,(\ref{param}) (blue solid line).}
\label{screeningm}
\end{figure}

The free energy obtained from lattice QCD calculations can be decomposed as,

\begin{eqnarray}
F(r,T)=F_c(r,T)+F_s(r,T),
\end{eqnarray}
where $F_c$ and $F_s$, respectively, denote Coulomb-like and the string terms and are expressed as

\begin{eqnarray}
F_c(r,T)&=&-\frac{\alpha}{r}[e^{-\mu r}+\mu r],\nonumber\\
F_s(r,T)&=&\frac{\sigma}{\mu}\bigg[\frac{\Gamma(1/4)}{2^{3/2}\Gamma(3/4)}-\frac{\sqrt{\mu r}}{2^{3/4}\Gamma(3/4)}K_{1/4}[(\mu r)^2]\bigg],\nonumber\\
\end{eqnarray}
with $\alpha=\pi/12$, $\sigma=0.445~{\rm GeV^2}$, and $\mu$ being the screening mass~\cite{Satz:2005hx}.
In Fig.~\ref{screeningm} we show $\mu$ extracted from lattice calculations \cite{Satz:2005hx,Kaczmarek:1900zz,Digal:2005ht} as a function of temperature. We
parametrize it separately below and above $T_c$ as follows:

\begin{eqnarray}
\frac{\mu}{\sqrt{\sigma}}&=&0.35+0.0034 \exp[(T/T_c)^2/0.22]\nonumber\\
&&~~~~~~~~~{\rm for~} T<T_c,\nonumber\\
\frac{\mu}{\sqrt{\sigma}}&=&0.45+0.5\bigg(\frac{T}{T_c}\bigg) \tanh\big[(T/T_c-0.93)/0.15\big] \nonumber\\
&&~~~~~~~~~{\rm for~} T>T_c.
\label{param}
\end{eqnarray}

The two separate parametrizations are smoothly matched at $T_c$, that is, with (almost) the same derivative.
Fig.~\ref{screeningm} shows that the screening mass rapidly increases near $T_c$.
If the free energy is interpreted as the heavy quark potential, this sudden increase brings about the lowering of the heavy quark potential,
which causes the quarkonia to melt at relatively low temperatures.

The internal energy is derived from thermodynamic relations as
\begin{eqnarray}
U(r,T)=F(r,T)+TS(r,T)\nonumber\\
=F(r,T)-T\frac{\partial F(r,T)}{\partial T}.
\end{eqnarray}

As mentioned above, the free energy drops sharply near $T_c$ and hence the entropy density defined as $-\partial F/\partial T$ rapidly increases.
As a result, if the internal energy is interpreted as the heavy quark potential,
a large potential wall is generated near $T_c$, causing the
quarkonia to be strongly bound and to dissolve only at higher temperatures.

The heavy quark potential is used in the Schr\"{o}dinger
equation of a charm and anti-charm pair as
\begin{eqnarray}
\bigg[2m_c-\frac{1}{m_c}\nabla^2+V(r,T)\bigg]\psi(r,T)=M \psi(r,T),\label{schrodinger1}
\end{eqnarray}
where $m_c=1.25$ GeV is the bare charm quark mass and $\psi(r,T)$
the charmonium wave function at temperature $T$.
Introducing the potential energy at infinity as
$V(r=\infty, T)$, Eq.\,(\ref{schrodinger1}) is modified
as~\cite{Karsch:1987pv,Satz:2005hx}
\begin{align}
\bigg[-\frac{\nabla^2}{m_c}+\widetilde{V}(r,T)\bigg]\psi(r,T)
=-\varepsilon\psi(r,T),
\label{schrodinger2}
\end{align}
where $\widetilde{V}(r,T)\equiv V(r,T)-V(r=\infty, T)$, which vanishes
at infinity, while $\varepsilon$ is the $J/\psi$ binding energy at temperature $T$,
\begin{eqnarray}
\varepsilon = 2m_c + V(r=\infty,T) - M.
\end{eqnarray}

The heavy quark potential in principle also has an imaginary part. Since it
however does neither much change the eigenenergy nor the eigenfunction of the Schr\"odinger equation, we ignore it
in the present study~\cite{Laine:2006ns,Brambilla:2008cx,Rothkopf:2011db}.
At vanishing binding energy, the eigenvalue $M$ of Eq.~(\ref{schrodinger1}) can be interpreted as the sum of the masses of two open heavy flavors, $D$ and $\bar{D}$ mesons in a
hadron gas or dressed charm and anticharm quarks in the QGP,

\begin{eqnarray}
m_D(T)\equiv \frac{M}{2} =m_c+\frac{1}{2}V(r=\infty,T).
\label{dmass}
\end{eqnarray}

In other words, if the charm and anticharm quarks that form a quarkonium state
are separated from each other by an infinite distance, the energy of the charm pair becomes the energy of two open heavy flavors.

From the above discussion, it is understood that the $D$ meson mass at finite temperature is closely related to the heavy quark
potential at infinity.
The asymptotic values of the free and internal energies are given by

\begin{eqnarray}
\lim_{r\rightarrow \infty} F(r,T)&=& \frac{\Gamma(1/4)}{2^{3/2}\Gamma(3/4)}\frac{\sigma}{\mu}-\alpha \mu, \nonumber\\
\lim_{r\rightarrow \infty} U(r,T)&=&\frac{\Gamma(1/4)}{2^{3/2}\Gamma(3/4)}\frac{\sigma}{\mu}-\alpha \mu\nonumber\\
&&+ T\frac{d\mu}{dT}\bigg[\frac{\Gamma(1/4)}{2^{3/2}\Gamma(3/4)}\frac{\sigma}{\mu^2}+\alpha\bigg].
\label{eqinfV}
\end{eqnarray}

\begin{figure}
\centerline{
\includegraphics[width=8.6 cm]{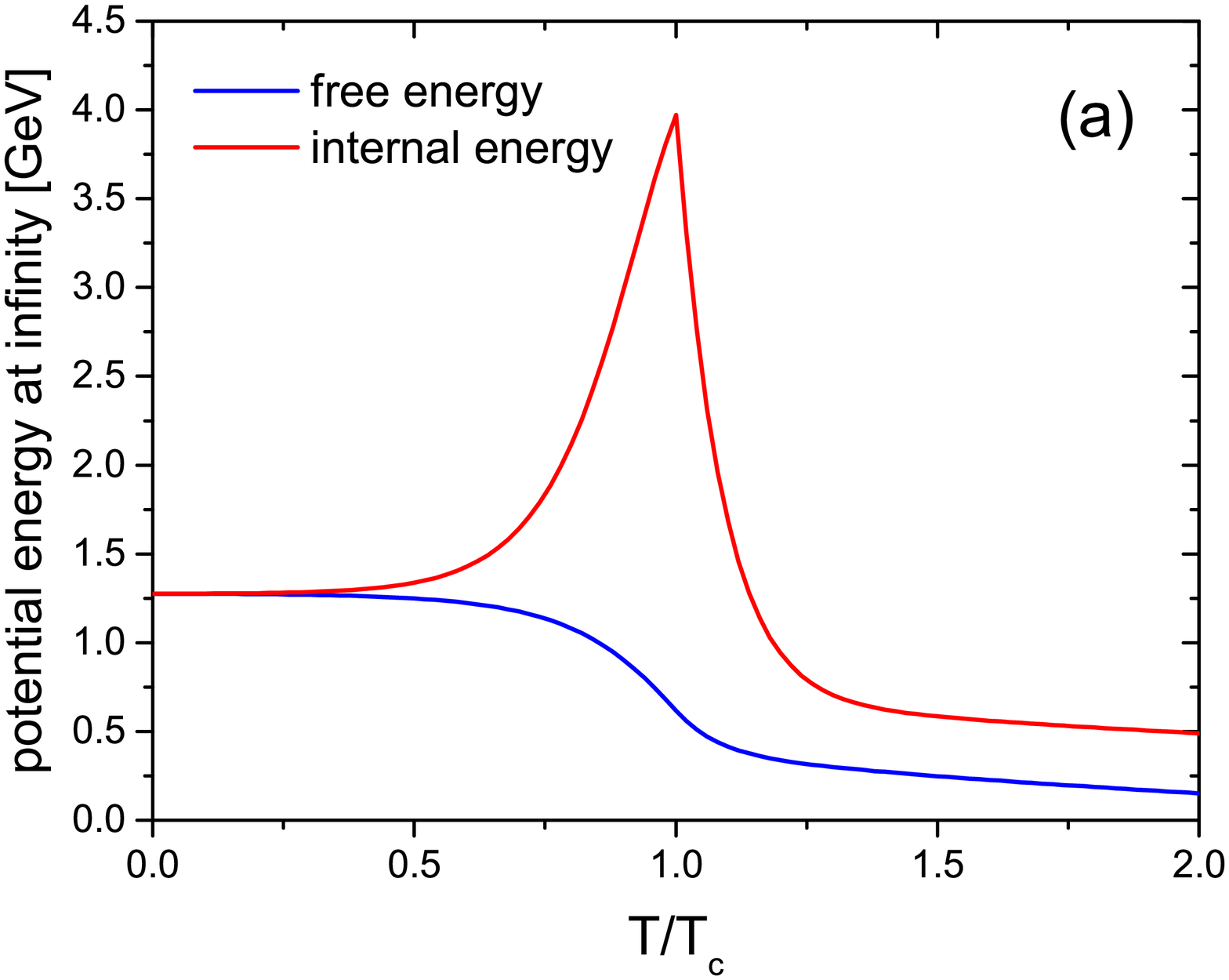}}
\centerline{
\includegraphics[width=8.6 cm]{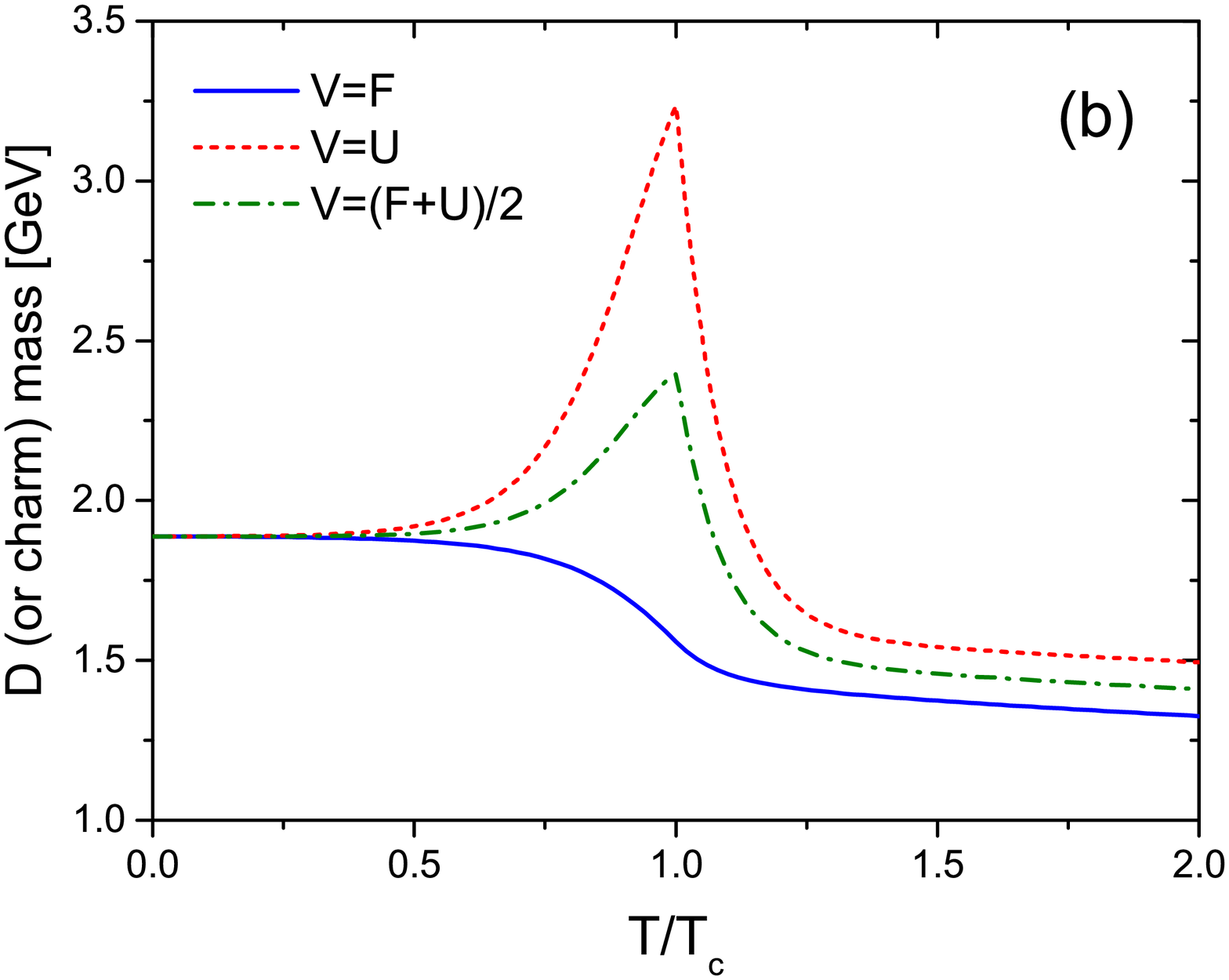}}
\caption{(a) The free energy and internal energy potentials at infinity and (b) the $D$ meson or dressed charm quark mass for several heavy quark potentials as functions of temperature.}
\label{infV}
\end{figure}

The upper panel of Fig.~\ref{infV} shows the free energy and internal energy potential values at infinity as functions of temperature, obtained from Eqs.~(\ref{param}) and (\ref{eqinfV}).
We note that the asymptotic values of the internal energy are very close to those of Refs.~\cite{Satz:2005hx,Digal:2005ht}.

The $D$ meson or dressed charm quark mass is shown in the lower panel of Fig.~\ref{infV} as a function of temperature for the free energy potential, the internal energy potential,
and the averaged potential.
One can see in this figure that the behavior of the $D$ meson mass is completely different, depending on the employed potential.
If the free energy is used as the potential, the $D$ meson mass decreases with increasing temperature, while the mass sharply increases up to 3.2 GeV near $T_c$ for the internal energy potential .
Even if one takes as the heavy quark potential the average of the free and internal energy, the D meson
mass increases up to 2.4 GeV around $T_c$.

\section{$D$ meson mass from QCD sum rules}\label{QSR}
In this section, we discuss the QCD sum rule approach used to compute the D meson mass at finite temperature.
As it is costumary in a QCD sum rule analysis \cite{Shifman:1978bx,Shifman:1978by},
we start with the correlator of an operator $j(x) = \bar{q}(x) i \gamma_5 c(x)$, which couples to
the $D$ meson state of interest,
\begin{eqnarray}
\Pi(q_0^2, \bm{q}^2, T) = i \int d^4x e^{iqx} \langle T[ j(x) j^{\dagger}(0) ] \rangle_{T}.
\label{correlator}
\end{eqnarray}
In this paper, we will assume that the $D$ is at rest, will hence set $\bm{q} = 0$ and
omit this variable for the rest of the discussion for simplicity of notation.
The spectral function $\rho(q_0^2, T) = 1/\pi \tanh(q_0/2T) \mathrm{Im} \Pi(q_0^2, T)=1/\pi  \mathrm{Im} \Pi^{\rm R}(q_0^2, T)$,
where the subscript R denotes the retarded correlation function, encodes the behavior of the $D$ meson
at finite density.
With the help of the analyticity of $\Pi^{\rm R}(q_0^2, T)$, the correlator in the deep Euclidean region ($q_0^2 \to - \infty$),
calculated through the time ordered correlator, can
be related to an integral over $\rho(q_0^2, T)$ via a dispersion relation.
Applying furthermore the Borel transform, one obtains the most
commonly used version of QCD sum rules, which reads
\begin{eqnarray}
\widehat{\Pi}(M^2, T) = \int_{0}^{\infty} ds e^{-s/M^2} \rho(s, T),
\label{sum_rules}
\end{eqnarray}
where $\widehat{\Pi}(M^2, T)$ stands for the Borel transformed correlator and $M$ is the so-called Borel mass.
For our purposes, we can identify the imaginary part of the time ordered correlator with the spectral density as the
thermal factor $\tanh\bigl[\sqrt{s}/(2T)\bigr]$ in the integrand on the right-hand side for the
temperature and energy values considered here ($\sqrt{s} \gtrsim m_D$ and $T \lesssim T_c$) is close to 1 and
can be safely neglected (see also the discussion in Ref.\,\cite{Buchheim:2018kss}).

As mentioned above, the left-hand side of Eq.\,(\ref{sum_rules}) is computed in the deep Euclidean region.
This means that one can rely on the operator product expansion (OPE), which is valid in this regime.
We will here not go into the details of this calculation, which have been discussed extensively in the literature (see
especially Ref.\,\cite{Zschocke:2011aa} for details), but only give the final result to be used in our analysis.
The leading order perturbative part of the OPE is given as
\begin{eqnarray}
\widehat{\Pi}^{\mathrm{pert}}(M^2, T) = \frac{1}{\pi} \int_{m^2_c}^{\infty} ds e^{-s/M^2} \mathrm{Im} \Pi^{\mathrm{pert}}(s, T)
\label{OPE_pert_1}
\end{eqnarray}
with
\begin{eqnarray}
\mathrm{Im} \Pi^{\mathrm{pert}}(s, T) = \frac{3}{8 \pi} s \Bigl(1 - \frac{m^2_c}{s}\Bigr)^2  \Bigl(1 + \frac{4}{3} \frac{\alpha_s}{\pi} R(m^2_c/s) \Bigr)
\label{OPE_pert_2}
\end{eqnarray}
and
\begin{eqnarray}
R(x) &=& \frac{9}{4} + \mathrm{Li}_2(x) + \ln (x) \ln(1-x) \nonumber\\
&&- \frac{3}{2} \ln \Bigl(\frac{1-x}{x}\Bigr) - \ln(1-x) + x \ln \Bigl(\frac{1-x}{x}\Bigr) \nonumber \\
&&- \frac{x}{1-x} \ln (x).
\label{OPE_pert_3}
\end{eqnarray}
This term does not depend on the temperature $T$. Next, the non-perturbative scalar condensate terms
up to dimension 5 are
\begin{eqnarray}
\widehat{\Pi}^{\langle \bar{q} q \rangle}(M^2, T) &=& -m_c \langle \bar{q} q \rangle_{T} e^{-m^2_c/M^2}, \\
\widehat{\Pi}^{\langle \frac{\alpha_s}{\pi} G^2 \rangle}(M^2, T) &=& \frac{1}{12} \Bigl \langle \frac{\alpha_s}{\pi} G^2 \Bigr \rangle_{T} e^{-m^2_c/M^2}, \\
\widehat{\Pi}^{\langle \bar{q} g \sigma G q \rangle}(M^2, T) &=& \frac{1}{2} \Bigl(\frac{m^3_c}{2M^4} - \frac{m_c}{M^2} \Bigr) \nonumber \\
&& \times \langle \bar{q} g \sigma G q \rangle_{T} e^{-m^2_c/M^2},
\label{OPE_non_pert_1}
\end{eqnarray}
while the non-perturbative non-scalar condensate terms are
\begin{eqnarray}
\widehat{\Pi}^{G_2}(M^2, T) &=& \frac{1}{4}  \Biggl[\Bigl(\frac{7}{6} \ln \frac{\mu^2 m^2_c}{M^4} + 2\gamma_{E} \Bigr) \Bigl(\frac{m^2_c}{M^2} - 1\Bigr) \nonumber \\
&& - 2 \frac{m^2_c}{M^2} \Biggr] G_2(T) e^{-m^2_c/M^2}, \\
\widehat{\Pi}^{F}(M^2, T) &=& \frac{3}{2} \Bigl(\frac{m^2_c}{M^2} - 1\Bigr) F(T) e^{-m^2_c/M^2}, \\
\widehat{\Pi}^{H}(M^2, T) &=&-3 \Bigl( \frac{m^3_c}{2M^4} - \frac{m_c}{M^2} \Bigr) H(T) e^{-m^2_c/M^2}.
\label{OPE_non_pert_2}
\end{eqnarray}
The sign in front of the $\gamma_{E}$ (the Euler-Mascheroni constant) term is different from the expression given
in Ref.\,\cite{Buchheim:2018kss}, which
can be traced back to a mistake/typo in Eq.\,(65) of Ref.\,\cite{Zschocke:2011aa} and
which is corrected here.
$G_2(T)$, $F(T)$ and $H(T)$ are related to non-scalar gluon and quark condensates as
\begin{eqnarray}
\Bigl \langle \mathcal{ST} \frac{\alpha_s}{\pi}  G^{a \mu}_{\alpha} G^{a \alpha \nu} \Bigr \rangle_T &=& G_2(T) \mathcal{ST}(u_{\mu} u_{\nu}), \nonumber \\
&=& G_2(T)(u_{\mu} u_{\nu} - \frac{1}{4} g_{\mu \nu} ), \\
\langle \mathcal{ST} \bar{q} \gamma_{\mu} i D_{\nu} q \rangle_T  &=& F(T) \mathcal{ST}(u_{\mu} u_{\nu}), \\
\langle \mathcal{ST} \bar{q} i D_{\mu} i D_{\nu} q \rangle_T &=& H(T) \mathcal{ST}(u_{\mu} u_{\nu}),
\label{non_scalar_cond}
\end{eqnarray}
where $\mathcal{ST}$ stands for the operation of making the following non-contracted Lorentz indices symmetric and
traceless.
$u_{\mu}$ is the four-velocity of the heat bath, which we assume to be at rest, hence $u_{\mu} =(1,0,0,0)$.
Higher order terms containing dimension 6 four quark condensates were computed in Ref.\,\cite{Buchheim:2014rpa}, but
turned out to be small. We therefore omit them in this work for simplicity.
In the QCD sum rule approach considered here, the temperature dependence is assumed to appear only in the QCD condensates and not in their
Wilson coefficients. Remembering that the OPE is realized through a division of scales, with low-energy components entering the
operators and high-energy contributions expressed as Wilson coefficients (with the boundary roughly at $\Lambda_{QCD}$),
it becomes clear that the above assumption is only valid as long as the temperature is low enough. Usually, one hence regards
this OPE to be applicable for temperatures up to and around $T_c$.

For the OPE input parameters at $T = 0$, we employ the values given in Table \ref{inputparameters}.
\begin{table}
\begin{center}
\caption{Vacuum input parameters, given at a renormalization scale of 2 GeV. Note that the $m_c$
used here is the pole mass.}
\label{inputparameters}
\begin{ruledtabular}
\begin{tabular}{ccc}
Input parameter & Value & Reference \\
\hline
$m_c$ & $1.67\,\mathrm{GeV}$ & \cite{Tanabashi:2018oca} \\
$\alpha_s$& 0.30 & \cite{Tanabashi:2018oca} \\
$\langle \bar{q}q \rangle_0$ & $(-0.272\,\mathrm{GeV})^3$ & \cite{Aoki:2019cca} \\
$\langle \frac{\alpha_s}{\pi} G^2 \rangle_{0}$ & $0.012\,\mathrm{GeV}^4$ & \cite{Shifman:1978bx,Shifman:1978by} \\
$\langle \bar{q}g \sigma G q \rangle_0/\langle \bar{q}q \rangle_0$ & $0.62\,\mathrm{GeV}^2$ & \cite{Belyaev:1982sa}
\end{tabular}
\end{ruledtabular}
\end{center}
\end{table}
Note that the pole mass used for the charm quark in the sum rule analysis is different from the value used
in the quark model calculation of the previous section, which
can be understood by considering the different renormalization point for the two cases.
The charm quark mass in the quark model is renormalized roughly at a typical momentum scale of the charm
quark in the studied system.
For QCD sum rules, one   estimates the relevant
two-point function in the deep-Euclidean region with non-perturbative effects taken into account through power corrections.
As was already noted a long time ago, for the Borel sum rules involving the heavy quarks  such as the charmonium
sum rules \cite{Bertlmann:1981he}, the pole mass value of $m_c = 1.67$ GeV is more suitable for this purpose as such a prescription tends to reduce the power corrections relative to the perturbative series, which
is also true for the D meson.  This is why we employ it here as well.
Our basic strategy in this paper is to fix the input charm quark masses such that a reasonably good
description of the D meson mass in vacuum is realized for both approaches and after that to obtain its
temperature dependence for fixed respective quark mass values.

To compute the strong coupling constant $\alpha_s$, the expression provided in the PDG \cite{Tanabashi:2018oca} with $N_f = 4$ and
$\Lambda^{(4)}_{\bar{MS}} = 292\,\mathrm{MeV}$ is used.
For $\langle \bar{q}g \sigma G q \rangle_0/\langle \bar{q}q \rangle_0$, we use the standard value of $0.8\,\mathrm{GeV}^2$,
renormalized at 1 GeV and run it to 2 GeV making use of its anomalous dimension \cite{Yang:1993bp}.
The temperature dependence of the condensates is obtained from different sources. Whenever possible, we have implemented
state-of-the-art lattice QCD results. Especially for the most important chiral condensate $\langle \bar{q}q \rangle_T$ (as well
as the less important $\langle \frac{\alpha_s}{\pi} G^2 \rangle_{T}$),
full QCD, continuum extrapolated results with $2 + 1$ flavors are available at the physical pion mass.
Specifically, we make use of the lattice data given in Ref.\,\cite{Borsanyi:2010bp} for $\langle \bar{q}q \rangle_T$ and those of Ref.\,\cite{Bazavov:2014pvz}
for $\langle \frac{\alpha_s}{\pi} G^2 \rangle_{T}$, according to the prescription discussed in Ref.\,\cite{Gubler:2018ctz}.
For $\langle \bar{q}g \sigma G q \rangle_T$, which so far has not been studied extensively on the lattice, we assume
$\langle \bar{q}g \sigma G q \rangle_T/\langle \bar{q}q \rangle_T$ to be constant and can therefore extract
the $\langle \bar{q}g \sigma G q \rangle_T/\langle \bar{q}q \rangle_T$ temperature dependence
directly from $\langle \bar{q}q \rangle_T$. The above assumption is supported by the lattice QCD calculation of Ref.\,\cite{Doi:2004jp}.
The non-scalar quark condensates encoded in $G_2(T)$, $F(T)$ and $H(T)$ are the least well known input parameters for the sum rules.
To estimate them, we apply the hadron resonance gas model, which assumes that the effect of finite temperature can
be described by a gas of non-interacting pions (and other light hadrons). This assumption should be approximately valid below $T_c$, where
hadronic degrees of freedom dominate, but becomes questionable around or above $T_c$, where quarks and gluons start to appear.
For concrete formulas and other details, we refer the interested reader to
Ref.\,\cite{Gubler:2018ctz}, and just mention here that we have included all small-mass pesudo-scalar particles, namely pions, kaons and the eta
in the actual calculation.
To obtain $G_2(T)$, one furthermore needs the value of the strong coupling constant at finite temperature, $\alpha_s(T)$.
To extract it, we use the two-loop perturbative running coupling, as given in Ref.\,\cite{Morita:2007hv} (see also Ref.\,\cite{Kaczmarek:2004gv}).

With all the OPE input determined, we next discuss how to analyze the sum rules of Eq.\,(\ref{sum_rules}). As it is standard, we assume the
spectral function to have the so-called ``pole + continuum'' form
(see, however, Ref.\,\cite{Gubler:2010cf} for an alternative approach),
\begin{eqnarray}
\rho(s, T) &=&
\lambda_D \delta(s - m^2_D) \nonumber \\
&& + \theta(s - s_{th}) \frac{1}{\pi} \mathrm{Im} \Pi^{\mathrm{pert}}(s, T),
\label{pole_cont}
\end{eqnarray}
which can be expected to be justified when the $D$ meson ground state peak dominates the spectral function at low energy and the perturbative
expression $\Pi^{\mathrm{pert}}(s, T)$ describes the spectral function above the threshold parameter $s_{th}$ reasonably well.
Some care is however needed when assessing the validity of the above assumption. While it may be sufficiently accurate in vacuum ($T = 0$), it will
eventually break down once the peak starts to dissolve into the continuum at some temperature above $T_c$.
Hence, analyses based on Eq.\,(\ref{pole_cont}) can only be trusted for temperatures up to and around $T_c$.

\begin{figure}
\centerline{
\includegraphics[width=8.6 cm]{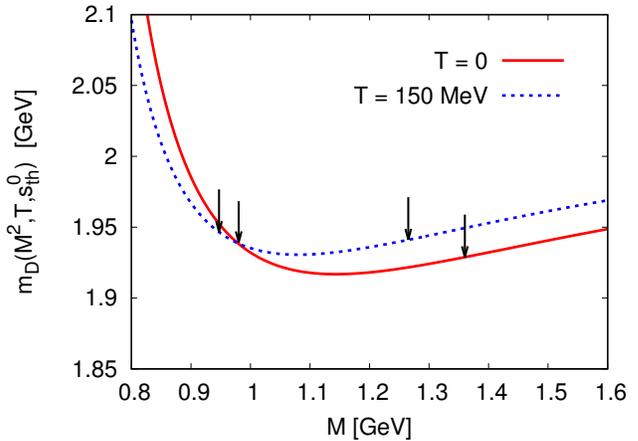}}
\caption{The $D$ meson masses computed from Eq.\,(\ref{sum_rules_4}) at $s_{th}^0$, which gives the maximum flatness, in the vacuum and at $T = 150$ MeV. The vertical
arrows indicate the locations of the Borel window boundaries, $M_{\mathrm{min}}$ and $M_{\mathrm{max}}$.}
\label{Borel_curve_comp}
\end{figure}

Substituting Eq.\,(\ref{pole_cont}) into Eq.\,(\ref{sum_rules}), we have
\begin{eqnarray}
\widehat{\Pi}(M^2, T) &=& \lambda_D e^{-m^2_D/M^2} \nonumber \\
&& + \frac{1}{\pi} \int_{s_{th}}^{\infty} ds e^{-s/M^2} \mathrm{Im} \Pi^{\mathrm{pert}}(s, T).
\label{sum_rules_2}
\end{eqnarray}
Therefore, we define
\begin{eqnarray}
&& \widetilde{\Pi}(M^2, T, s_{th}) \equiv \nonumber \\
&& \widehat{\Pi}(M^2, T) - \frac{1}{\pi} \int_{s_{th}}^{\infty} ds e^{-s/M^2} \mathrm{Im} \Pi^{\mathrm{pert}}(s, T),
\label{sum_rules_3}
\end{eqnarray}
and can compute the $D$ meson mass as
\begin{eqnarray}
m_D(M^2, T, s_{th}) = \sqrt{-\frac{ \partial \widetilde{\Pi}(M^2, T, s_{th})/\partial(1/M^2)}{\widetilde{\Pi}(M^2, T, s_{th})}},
\label{sum_rules_4}
\end{eqnarray}
and subsequently the residue $\lambda_D$ as
\begin{eqnarray}
\lambda_D (M^2, T, s_{th}) = \widetilde{\Pi}(M^2, T, s_{th}) e^{m^2_D(M^2, T, s_{th})/M^2}.
\label{sum_rules_5}
\end{eqnarray}
We, however, still need to fix $M$ and $s_{th}$, which are artificial parameters of the sum rule approach.
Starting with $M$, we define a so-called Borel window, for which the approximations used in the sum rule analysis can be expected to be
valid. The lower boundary of the Borel window, $M_{\mathrm{min}}$, is determined from the condition of sufficient convergence of the OPE,
namely
\begin{eqnarray}
\frac{\widehat{\Pi}^{\mathrm{OPE}}_{\mathrm{dim\,5\,terms}}(M^2, T)}{ \widehat{\Pi}^{\mathrm{OPE}}_{\mathrm{all\,terms}}(M^2, T) } < 0.1.
\label{Borel_window_1}
\end{eqnarray}
Here, $\widehat{\Pi}^{\mathrm{OPE}}_{\mathrm{dim\,5\,terms}}(M^2, T)$ is the sum of the terms of condensates with the highest dimension considered
in this work, which are $\widehat{\Pi}^{\langle \bar{q} g \sigma G q \rangle}(M^2, T)$ and $\widehat{\Pi}^{e_2}(M^2, T)$. For the upper boundary, $M_{\mathrm{max}}$,
we employ the condition that the pole contribution in the integral of Eq.\,(\ref{sum_rules}) should dominate the sum rule. Specifically, we have
\begin{eqnarray}
\frac{  \int_{0}^{s_{th}} ds e^{-s/M^2} \rho(s, T)  }{ \int_{0}^{\infty} ds e^{-s/M^2} \rho(s, T) }
= \frac{ \widetilde{\Pi}(M^2, T, s_{th}) }{ \widehat{\Pi}(M^2, T) } > 0.5.
\label{Borel_window_2}
\end{eqnarray}
With the range of $M$ determined as $M_{\mathrm{min}} < M < M_{\mathrm{max}}$, one can compute averages of $m_D(M^2, T, s_{th})$ and $\lambda_D (M^2, T, s_{th})$.
We have
\begin{eqnarray}
\overline{m}_D(T, s_{th}) &=& \frac{1}{M_{\mathrm{max}} - M_{\mathrm{min}}} \nonumber \\
&& \times \int_{M_{\mathrm{min}}}^{M_{\mathrm{max}}} dM m_D(M^2, T, s_{th}), \\
\overline{\lambda}_D (T, s_{th}) &=& \frac{1}{M_{\mathrm{max}} - M_{\mathrm{min}}} \nonumber \\
&& \times \int_{M_{\mathrm{min}}}^{M_{\mathrm{max}}} dM \lambda_D (M^2, T, s_{th}).
\label{averages}
\end{eqnarray}
Finally, $s_{th}$ is determined as $s_{th}^0$, which minimizes the function
\begin{eqnarray}
&& d(T, s_{th}) =  \frac{1}{M_{\mathrm{max}} - M_{\mathrm{min}}} \nonumber \\
&& \times \int_{M_{\mathrm{min}}}^{M_{\mathrm{max}}} dM \Bigl[ m_D(M^2, T, s_{th}) - \overline{m}_D(T, s_{th}) \Bigr]^2.
\label{determine_s_th}
\end{eqnarray}
This ensures that, within the Borel window, the dependence of $m_D(M^2, T, s_{th})$ on $M$ is as small as possible.
For illustration, we show in Fig.\,\ref{Borel_curve_comp} $m_D(M^2, T, s_{th}^0)$ for $T = 0$ and $T = 150$ MeV as a function of $M$. The respective positions of
$M_{\mathrm{min}}$ and $M_{\mathrm{max}}$ are indicated as vertical arrows.

We have carried out two different analyses, one in which we let the Borel window depend on $T$ (shown in Fig.\,\ref{Borel_curve_comp}) and
another one in which we keep it fixed at its vacuum location.
\begin{figure}[h!]
\centerline{
\includegraphics[width=8.6 cm]{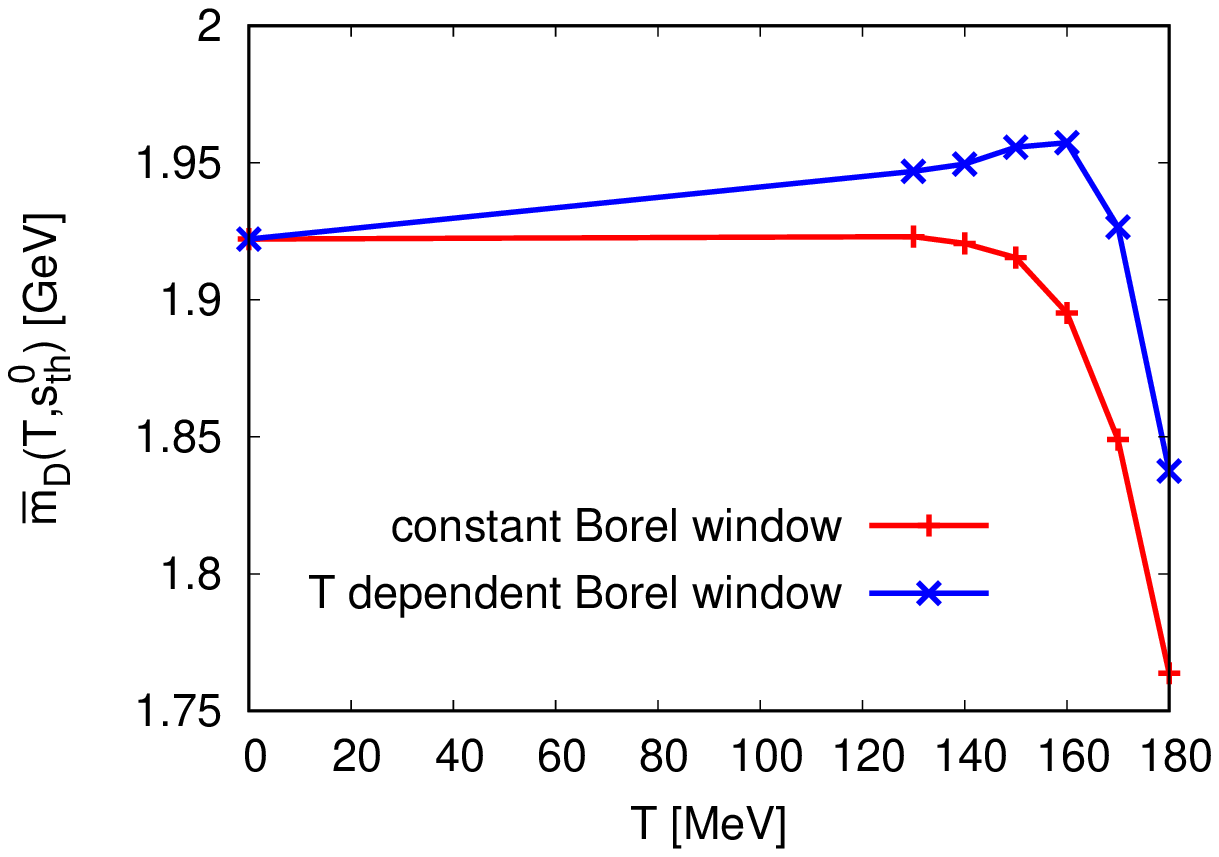}} 
\centerline{
\includegraphics[width=8.6 cm]{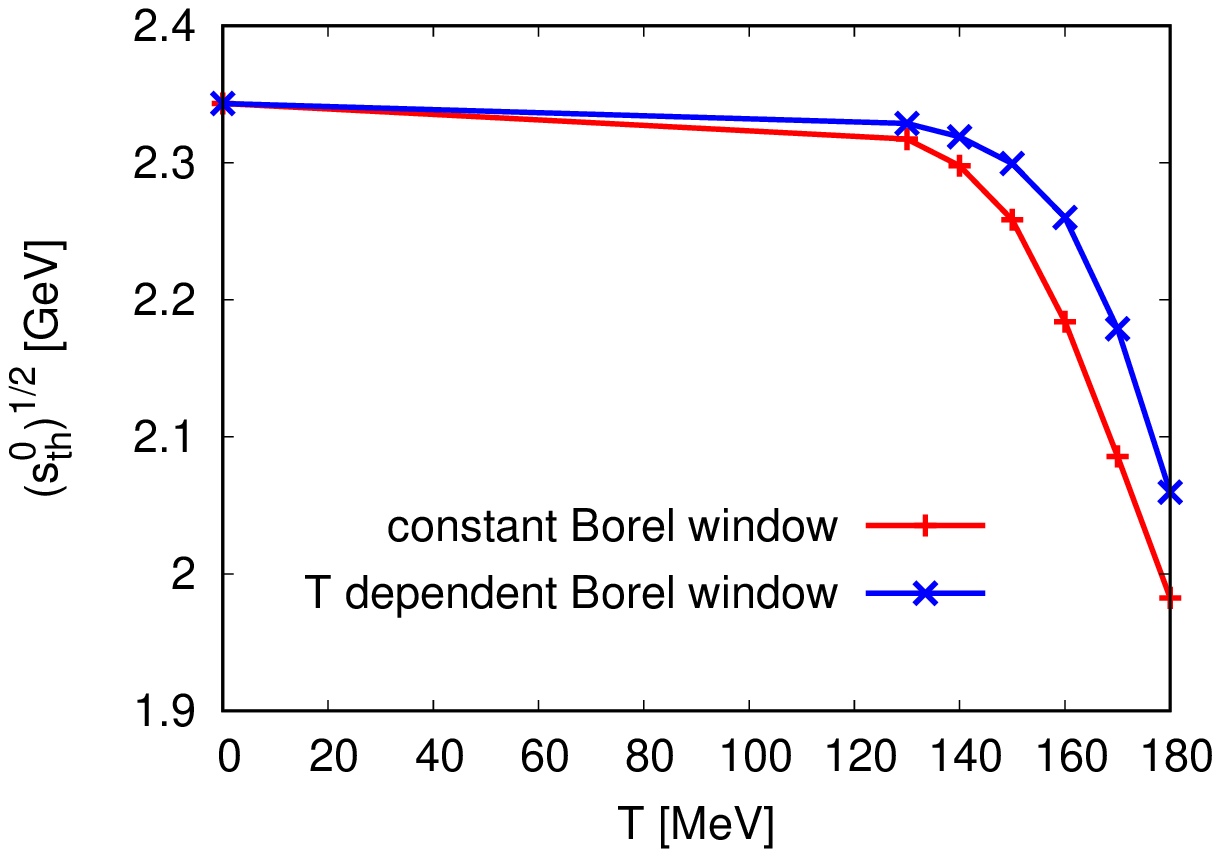}} 
\centerline{
\includegraphics[width=8.6 cm]{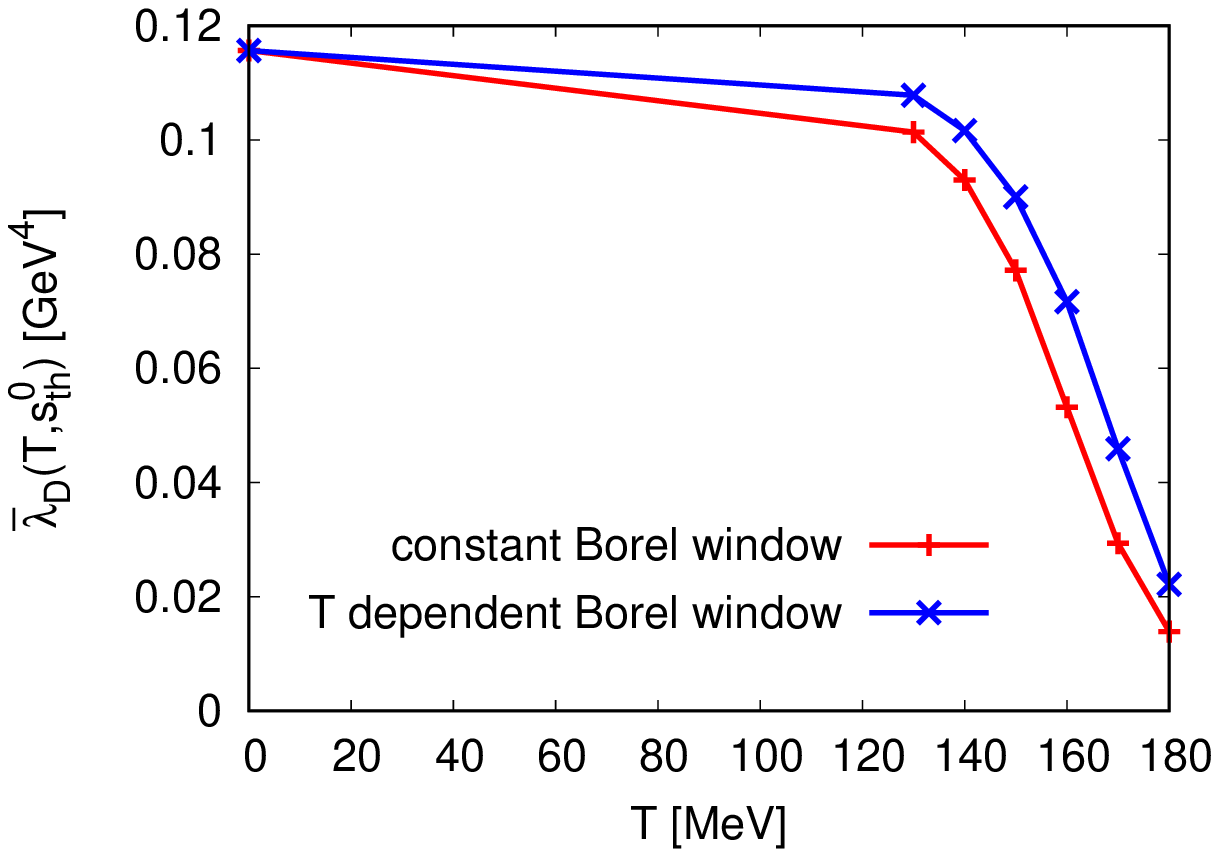}}
\caption{The values of $\overline{m}_D(T, s_{th}^0)$, $\sqrt{s_{th}^0}$, and $\overline{\lambda}_D (T, s_{th}^0)$ as a function of temperature $T$. The red curve (horizontal crosses) shows
the analysis results when the Borel window is determined at $T=0$ and kept fixed with increasing temperature. The blue curve (diagonal crosses) gives the results
for a temperature dependent Borel window.}
\label{Borel_window_comp}
\end{figure}
The two versions are shown in Fig.\,\ref{Borel_window_comp}, where it is seen in the top plot that
the $D$ meson mass increases or remains roughly constant below and around $T_c \simeq 155$ MeV \cite{Borsanyi:2010bp}
and starts to drop for higher temperatures.
In the middle plot of Fig.\,\ref{Borel_window_comp}, we show the temperature dependence of the threshold parameter $\sqrt{s_{th}^0}$,
which lies about 0.4 GeV above the mass value and behaves roughly in parallel with it as the temperature increases.
The residue $\lambda_{D}$, shown in the bottom plot of Fig.\,\ref{Borel_window_comp}, decreases monotonically with increasing temperature and
approaches zero towards $T = 180$ MeV.
Comparing the two analyses, it is observed that only the $T$ dependent Borel window case leads to an increase of the $D$ meson mass.
It also gives a slightly slower
decrease of the residue $\lambda_D$.

It is worth thinking about a physically intuitive picture for the above results.
First, let us consider the potential increase of the $D$ meson mass below and around $T_c$, which seems to be consistent with the reconstructed spectral functions given in
the recent lattice QCD study of Ref.\,\cite{Kelly:2018hsi}, the lattice data quality however not being good enough to draw any definite conclusions.
The most natural explanation for this behavior can be given by the quark model picture proposed in Ref.\,\cite{Park:2016xrw}, where the increase of
the $D$ meson mass was related to the decreasing constituent quark mass caused by the partial restoration of chiral symmetry. Due to the decreasing
quark mass, its wave function spreads out to larger distances and therefore feels the linear confining potential, leading to an overall increase of the
$D$ meson mass. As in the present study the chiral symmetry gets partially restored due to the decrease of the chiral condensate, the same quark
model picture provides a natural explanation for an increasing $D$ meson mass behavior around $T_c$.
Next, we examine the
decreasing $D$ meson mass at temperatures around and above $T_c$. To consider this behavior, one should take into
account the decreasing residue value $\lambda_{D}$, shown in the lower plot of Fig.\,\ref{Borel_window_comp}. This decrease means that the contribution
of the ground state peak to the spectral function is reduced, which can be interpreted as the gradual melting of the $D$ meson into the
continuum, which is expected to happen at some (unknown) temperature above $T_c$. Therefore, if the spectral function at $T > T_c$ is gradually
changed from a narrow peak to a continuum, the ``pole + continuum'' assumption of Eq.\,(\ref{pole_cont}) breaks down and the notion of a
$D$ meson mass loses any meaning.
It should however be kept in mind that
both the division of scales of the OPE and the estimation of some of the non-scalar QCD condensates become questionable above $T_c$.
The physical interpretation of the drop shown in the upper plot of Fig.\,\ref{Borel_window_comp} for $T > T_c$ hence
is not completely clear because of the limitations of the QCD sum rule method at such temperatures, but is likely related to the melting of the $D$ meson state
into the continuum.
In the next section, we will discuss further physical interpretations of the QCD sum rule results in terms of the quark model approach considered in the previous section.


\section{Matching the heavy quark potential to QCD sum rules}\label{comparison}

\begin{figure}[t!]
\centerline{
\includegraphics[width=8.6 cm]{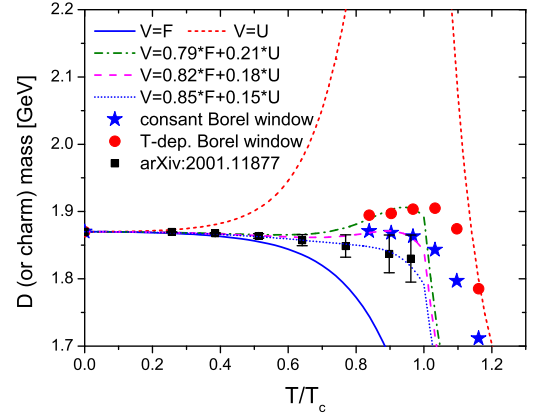}}
\caption{The $D$ meson or dressed charm quark mass for several heavy quark potentials as functions of temperature in comparison with the QCD sum rule results.
$T_c$ is taken to be 155 MeV. 
For the results from Ref.~\cite{Montana:2020lfi}, effects from spectral widths are shown as error bars.
All results are shifted to have the common physical $D$ meson mass in vacuum ($T=$0).}
\label{philipp-juan}
\end{figure}

Let us now study the consequences of requiring that the temperature dependence of the D meson mass
obtained from the potential model given in Eq.~(\ref{dmass}) is consistent with that from QCD sum rules.
As can be seen in Fig.\,\ref{philipp-juan}, assuming a constant or temperature dependent Borel window requires that
the heavy quark potential should have a nontrivial contribution from the internal energy in addition to the dominant free energy part.
The discrepancy above $T_c$ should not be taken too seriously as the $D$ meson mass calculations are not reliable at these temperatures, as mentioned in the previous section.
Also shown in the figure are the recent results from Ref.\,\cite{Montana:2020lfi} where the $D$ meson mass was calculated at finite temperature in the imaginary-time formalism
by using a hadronic effective field theory with chiral and heavy-quark spin symmetries.
Although the central value for the $D$ meson mass  slightly decreases as the temperature approaches the critical temperature,
the difference compared with the QCD sum rule result does not qualitatively change our discussion of the heavy quark potential of this section.
We find that  fitting the results in the potential model  to the $D$ meson mass obtained from QCD sum rules with either a constant or temperature dependent Borel window requires the
heavy quark potential to be composed of 18\% and 21\% of internal energy potentials, respectively, with the remaining contribution coming from the free energy potential.
The results from the hadronic effective model are fitted with 15\% of the internal energy, if one fits the potential to the central pole mass.

On the other hand, a previous determination of the heavy quark potential obtained by comparing the potential model result for the $J/\psi$ to a QCD sum rule calculation,
led to the conclusion that the heavy quark potential should be dominated by the free energy~\cite{Lee:2013dca}.
The two findings are, however, not necessarily inconsistent, as combining them, one can
naturally conclude that the effective heavy quark potential  is dominated by the free energy at short distance, while at larger separation, it will have  a nontrivial fraction of internal energy.
This is in line with the interpretation given in Ref.~\cite{Satz:2015jsa}. For short distances, the interquark potential should be affected by the gluon dynamics
exchanged between the two heavy quarks, which is encoded in the free energy.  On the other hand, as the separation between the heavy quarks becomes large,
the polarization of the heavy quark itself becomes important, which is related to the D meson, but at the same time introduces additional entropy contribution included in the internal energy.
The same effect can be simulated with using just the free energy potential but a larger effective heavy quark mass at larger distance.  The introduction of additional internal energy contribution
to the potential energy will become important when one calculates the binding energy and wave function of the excited states which are of larger size compared to the ground state.

\begin{figure}[h!]
\centerline{
\includegraphics[width=8.6 cm]{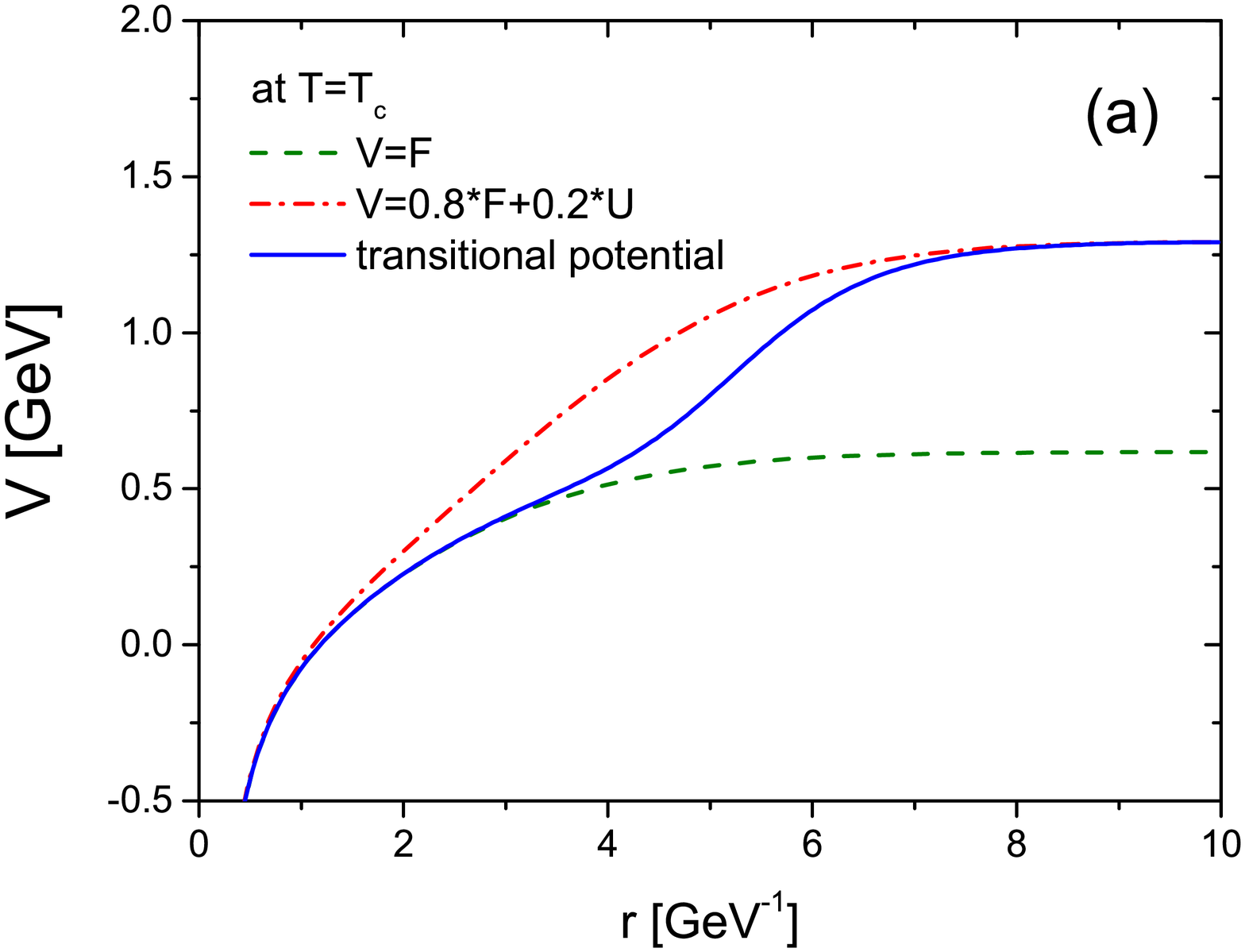}}
\centerline{
\includegraphics[width=8.6 cm]{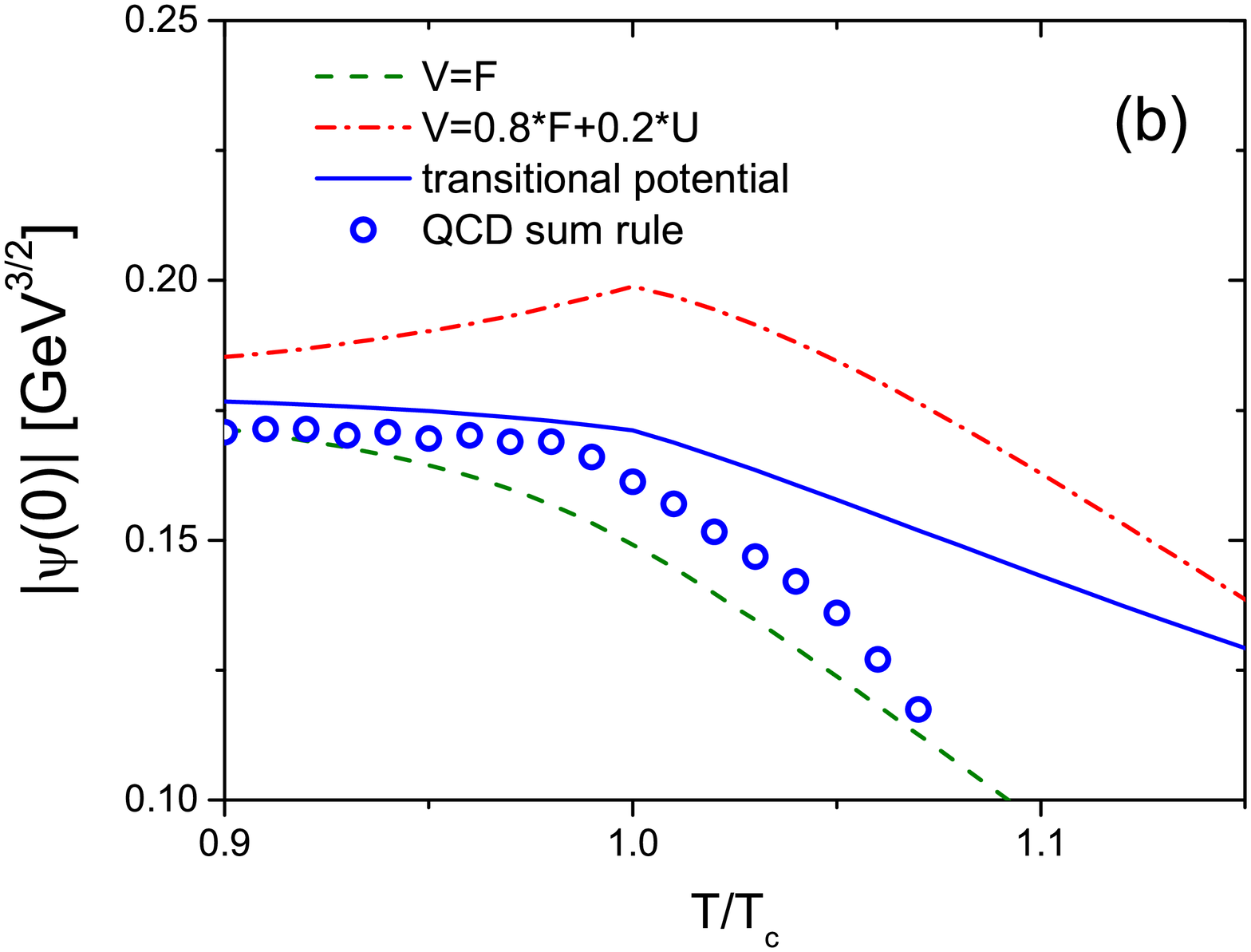}}
\centerline{
\includegraphics[width=8.6 cm]{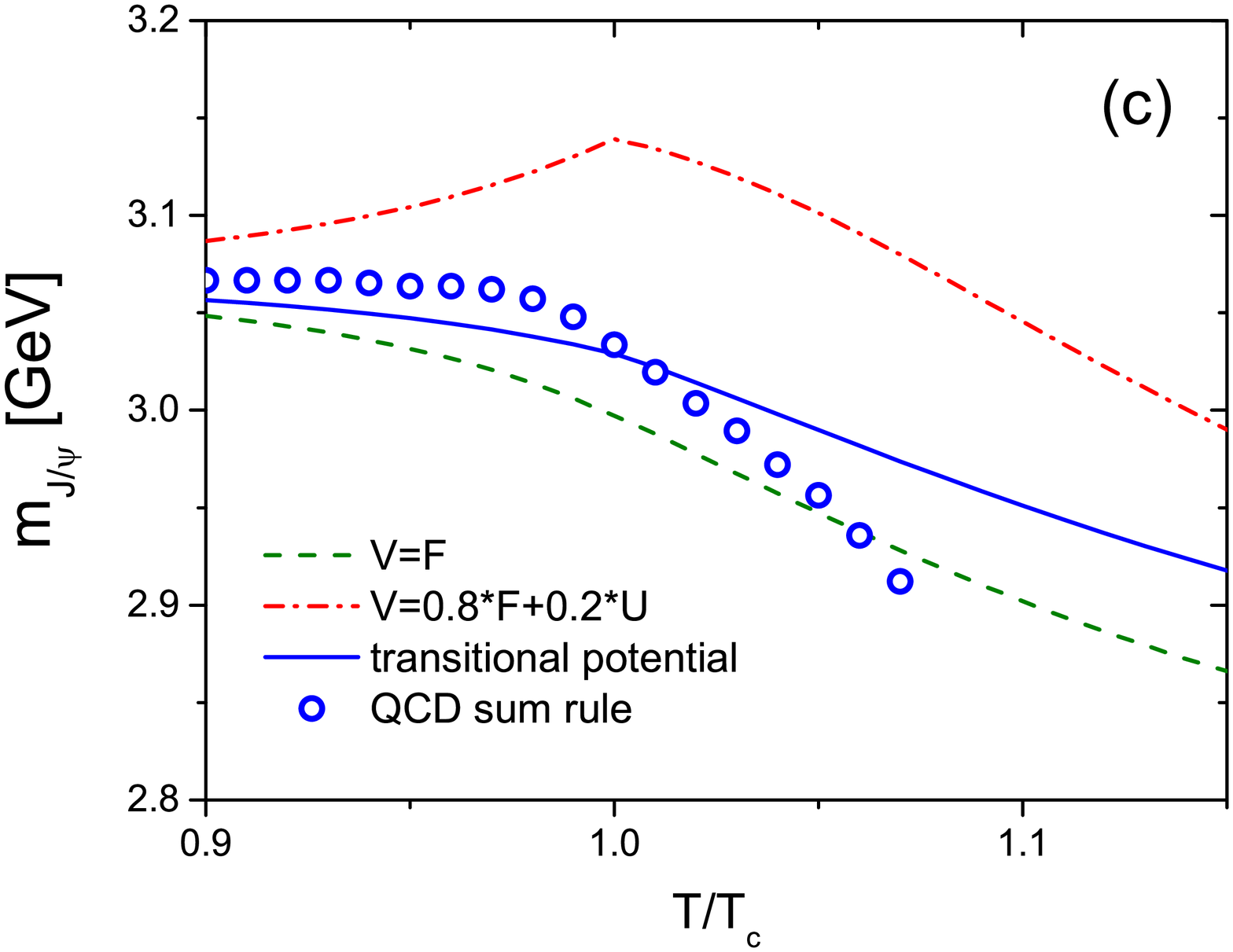}}
\caption{(a) Three different heavy quark potential energies, (b) the residue and (c) the mass of $J/\psi$ as a function of temperature for each heavy quark potential energy,
compared with the results from QCD sum rules~\cite{Lee:2013dca}.}
\label{jpsi}
\end{figure}

As an illustration we introduce a heavy quark potential, which starts with the free energy at short distance and then smoothly transits
to a combinational potential energy composed of 80\% of the free energy and 20\% of the internal energy with increasing interquark  distance, as shown in Fig.~\ref{jpsi}(a).
In Fig.~\ref{jpsi}(b) and (c), we show the result  for the
wavefunction at the origin and the $J/\psi$ mass obtained by solving the
Schr\"odinger equation of Eq.\,(\ref{schrodinger2})
by using three different potentials: the free energy potential, the combinational potential, and the transitional potential.
Also shown in the figures are the results obtained from QCD sum rules \cite{Lee:2013dca}.
The results for the free energy are slightly different from those reported in Ref.~\cite{Lee:2013dca} because of the parametrization of Eq.~(\ref{param}).
First, one notes that results obtained by using the
combinational potential energy with 20\% of the internal energy deviates from the temperature dependence obtained from the QCD sum rule approach.
However,
comparing the results between the free energy and the transitional potential, shows that the residue and mass of the $J/\psi$ from the latter better reproduces
the QCD sum rule results  at least up to the critical temperature where the $D$ meson mass is investigated in this study.

Furthermore, the quoted QCD sum rule result is obtained within the zero width approximation.
The mass shift required by the changes in the OPE can in general be accommodated by an increase in the width without any change in the mass as was first pointed out in Ref.\,\cite{Leupold:1997dg}.
This is so because the effects of a negative mass shift in the spectral density in the Borel transformed dispersion  relation
 can be approximated by an increase in the width as such a change also increases the contribution in the spectral density at the lower energy region.
Results of detailed past sum rule analyses show that the mass shift and width increase 
can be approximated by $\Delta \Gamma = {\rm Constant}+ a \Delta m $, where the proportionality coefficient $a$ is about 1.5 for the  $J/\psi$ sum rule \cite{Morita:2007hv} and 2.5 for the light vector meson sum rules \cite{Leupold:1997dg}. The ``Constant" here depends on the sum rule considered  and the temperature.
That means that if we allow for an additional width increase in the sum rule, an additional positive mass change should be added.
As the D meson lies roughly in the middle of the $J/\psi$ and the light vector mesons,
we can in the current study take the proportionality coefficient $a$ to be around 2.  Then, the width increase of 30-70 MeV
at T=150 MeV \cite{Fuchs:2004fh,Cleven:2017fun} for the D meson will inevitably lead  an additional mass increase of 15-35 MeV.
Such an additional mass increase is certainly non-negligible but  within the uncertainty in the sum rule analysis estimated by the difference in the results for the mass change obtained with a temperature independent and dependent Borel window shown respectively as blue stars and red circles in Fig.\,\ref{philipp-juan}.

In conclusion, our comparison suggests the heavy quark potential to be composed purely of the free energy at short distance
while at larger separation it has  a nontrivial contribution from the internal energy. This configuration can reproduce both the $J/\psi$ and the $D$ meson masses
at finite temperature obtained from
QCD sum rules up to and around the critical temperature.

\section{Summary and Conclusions}\label{conclusion}

We have
in this paper studied the heavy quark - anti-quark potential at finite temperature, which is crucial for
understanding the quarkonium production in relativistic heavy-ion collisions, through the behavior of the $D$ meson mass obtained by using QCD sum rules.
Based on the observation that the total energy of a heavy quark bound state pair
at infinity can be interpreted as twice the D meson mass at
finite temperature, it becomes possible to compute the heavy quark potential at a large distance from the thermal behavior of D mesons.
Our QCD sum rule results for the D meson at finite temperature suggest that the heavy quark potential at large distance should be composed dominantly of the
free energy but with a contribution of the internal energy of about 20\%.
Combined with a similar comparison
for the $J/\psi$ channel at finite temperature, we conclude that the heavy
quark potential should have a different fraction of internal energy as a function of the interquark distance.
Summarizing the result, we find that to reproduce the behavior of both the $J/\psi$ and
$D$ meson from QCD sum rules without contradiction at least up to and around $T_c$, the heavy quark potential should be composed of the free energy at
short distance, while attaining
a nontrivial  fraction of the internal energy at larger distances so that at infinity the fraction becomes about 20\%.

\section*{Acknowledgements}
The authors acknowledge useful discussions with E.
Bratkovskaya and J. M. Torres-Rincon.
This work was supported by Samsung Science and Technology Foundation under Project No. SSTF-BA1901-04.
P.G. is supported by the Grant-in-Aid for Early-Carreer Scientists (JSPS KAKENHI Grant No. JP18K13542), 
Grant-in-Aid for Scientific Research (C) (JSPS KAKENHI Grant No. JP20K03940) and
the Leading Initiative for Excellent Young Researchers (LEADER) of the Japan Society for the Promotion of
Science (JSPS).

\end{document}